\documentclass[journal,twoside,web]{ieeecolor}
\usepackage{generic}
\usepackage{cite}
\usepackage{amsmath,amssymb,amsfonts}
\usepackage{algorithmic}
\usepackage{graphicx}
\usepackage{textcomp}
\usepackage{textcomp}
\usepackage{array}
\usepackage{caption}
\usepackage[ruled,vlined]{algorithm2e}
\usepackage{booktabs}
\usepackage[caption=false,font=normalsize,labelfont=sf,textfont=sf]{subfig}

\usepackage{tikz}  
\tikzstyle{startstop} = [rectangle, rounded corners, minimum width=3cm, minimum height=1cm,text centered, draw=black, fill=red!30]
\tikzstyle{io} = [trapezium, trapezium left angle=70, trapezium right angle=110, minimum width=3cm, minimum height=1cm, text centered, draw=black, fill=blue!30]
\tikzstyle{process} = [rectangle, minimum width=3cm, minimum height=1cm, text centered, draw=black, fill=orange!30]
\tikzstyle{decision} = [diamond, minimum width=3cm, minimum height=1cm, text centered, draw=black, fill=green!30]
\tikzstyle{Initialization} = [rectangle, rounded corners, minimum width=3cm, minimum height=1cm, draw=black, fill=red!30]
\tikzstyle{process} = [rectangle, minimum width=3cm, minimum height=1cm, text centered, draw=black, fill=orange!30]
\tikzstyle{arrow} = [thick,->,>=stealth]

\def\BibTeX{{\rm B\kern-.05em{\sc i\kern-.025em b}\kern-.08em
    T\kern-.1667em\lower.7ex\hbox{E}\kern-.125emX}}
\begin{document}
\title{Distant Domain Transfer Learning for Medical Imaging}
\author{Shuteng~Niu, Meryl Liu, Yongxin~Liu, Jian Wang, and~Houbing~Song,~\IEEEmembership{Senior~Member,~IEEE}
}

\markboth{IEEE Journal of Biomedical Health Informatics, Vol. 00, No. 0, Month 2020}
{First A. Author \MakeLowercase{\textit{et al.}}: Bare Demo of IEEEtai.cls for IEEE Journals of Journal of Biomedical Health Informatics}

\maketitle

\begin{abstract}
Medical image processing is one of the most important topics in the field of the Internet of Medical Things (IoMT). Recently, deep learning methods have carried out state-of-the-art performances on medical image tasks. However, conventional deep learning have two main drawbacks: 1) insufficient training data and 2) the domain mismatch between the training data and the testing data. In this paper, we propose a distant domain transfer learning (DDTL) method for medical image classification. Moreover, we apply our methods to a recent issue (Coronavirus diagnose). Several current studies indicate that lung Computed Tomography (CT) images can be used for a fast and accurate COVID-19 diagnosis. However, the well-labeled training data cannot be easily accessed due to the novelty of the disease and a number of privacy policies. Moreover, the proposed method has two components: Reduced-size Unet Segmentation model and Distant Feature Fusion (DFF) classification model. It is related to a not well-investigated but important transfer learning problem, termed Distant Domain Transfer Learning (DDTL). DDTL aims to make efficient transfers even when the domains or the tasks are entirely different. In this study, we develop a DDTL model for COVID-19 diagnose using unlabeled Office-31, Catech-256, and chest X-ray image data sets as the source data, and a small set of COVID-19 lung CT as the target data. The main contributions of this study: 1) the proposed method benefits from unlabeled data collected from distant domains which can be easily accessed, 2) it can effectively handle the distribution shift between the training data and the testing data, 3) it has achieved 96\% classification accuracy, which is 13\% higher classification accuracy than "non-transfer" algorithms, and 8\% higher than existing transfer and distant transfer algorithms.
\end{abstract}


\section{Introduction}
\label{sec:introduction}

\begin{figure*}
	\centering
	
		\includegraphics[scale=.4]{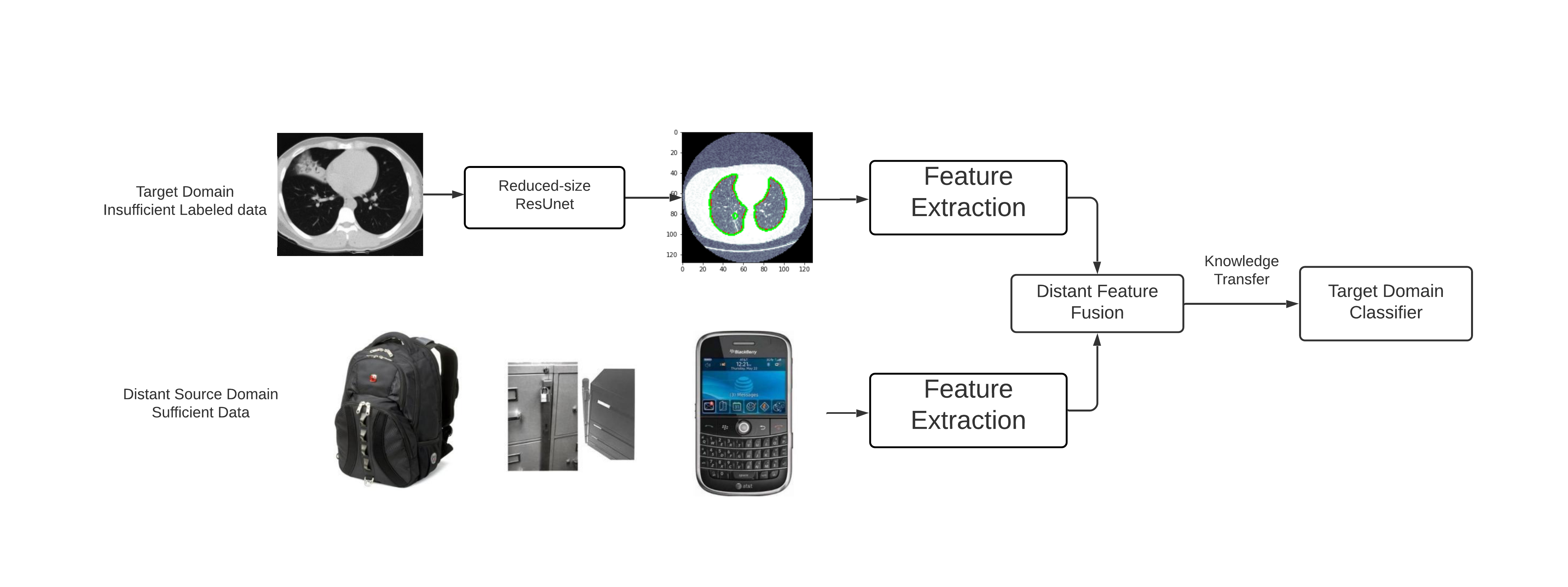}
	\caption{Architecture Overview of Distant Feature Fusion Model}
	\label{intro}
\end{figure*}

Recently, with state-of-art performance, deep learning methods have dominated the field of image processing. However, deep learning methods require a massive amount of well-labeled training data, and the majority of deep leaning methods are sensitive to the domain shift. Therefore, transfer learning (TL) has been introduced to deal with those two issues. In this paper, we propose a novel medical image classification algorithm. Moreover, we implement the proposed algorithm on a COVID-19 diagnose application. Generally, medical image data sets are difficult to access due the rarity of diseases and privacy policies. Moreover, it is difficult to manually collect a massive amount of high-quality labeled lung CT scans associated with of COVID-19. Thus, the performance of machine learning models can be unsatisfying with insufficient training data. To overcome this obstacle, artificial and synthetic data can be used to expand the volume of the data. However, these methods cannot handle the performance degradation caused by the distribution mismatch between the training data and the testing data. Therefore, transfer learning is considered to be one of the most effective ways to solve both problems at the same time. Theoretically, transfer learning algorithms aim to develop robust target models by using only a small set of target training data and transferring knowledge learned from other domains and tasks. Previously, \cite{sun2016deep} proposed an adaptation layer with domain distance measurements to transfer knowledge between deep neural networks. In general, conventional transfer learning algorithms assume that the source domains and the targets share a certain amount of information. However, this assumption does not always hold in many real-world applications, such as medical image processing \cite{shin2016deep, cheplygina2019not}, rare species detection \cite{taroni2019multiplier} and recommendation systems \cite{pan2010transfer,pan2011transfer}. Moreover, transferring between two loosely related domains usually causes negative transfer \cite{pan2009survey}, meaning that the knowledge transfer starts hurting the performance on the task in the target domain, producing worse performance than non-transfer models. For instance, building a dog classification model by directly transferring knowledge from a car classification model would likely to lead to negative transfer due to the weak connection between the two domains. Therefore, it is not always feasible to apply transfer learning to areas where we cannot easily obtain enough source domain data related to the target domain. COVID-19 diagnosis based on lung CT is a typical example where we cannot easily find related source data for training, so conventional transfer learning can lead to negative transfer. 

In this paper, we develop a lung CT scan-based COVID-19 classification framework by studying a challenging problem, distant domain transfer learning (DDTL), which aims to deal with the shortcomings of traditional machine learning and conventional transfer learning. As shown in Fig. \ref{intro}, the proposed framework contains two parts: semantic segmentation and distant feature fusion. It can perform knowledge transfer between distant domains which are seemingly unrelated. Moreover, DDTL \cite{tan2015transitive} is a newly introduced transfer learning method that mainly aims to address the issue of negative transfer caused by loose relations of the source domains and the target domains. In other words, it allows us to safely and effectively perform the knowledge transfer when the source domains and target domains only share a very weak connection. Unlike conventional TL methods, the proposed DDTL algorithm benefits from fusing distant features extracted from distant domains. Moreover, the inspiration for DDTL lies in the ability of human beings to learn new things by building on knowledge acquired from several seemingly independent things. For example, a human who knows birds and airplanes can recognize a rocket even without seeing any rockets previously. Therefore, DDTL dramatically extends the use of transfer learning to more areas, and applications where do not always have adequate related source data. As mentioned previously, there is no easy access to a sufficient amount of well-labeled lung CT images of COVID-19 because of the labeling difficulty, and patient privacy policies. Therefore, we consider this task as a DDTL problem that can benefit from distant but more accessible domains. For instance, we use three open-source image data sets as source domain data sets to develop a robust COVID-19 classification method based on lung CT images.

Moreover, unlike most existing methods, we introduce a feature-based algorithm with a novel distant feature fusion process. Previously, there are few proposed distant transfer algorithms \cite{tan2015transitive,tan2017distant}, but most of them are task-specific and lack stability in performance. Inspired by an instance-based method \cite{tan2017distant} and multi-task learning \cite{zhang2014facial}, we build a DDTL algorithm to solve COVID-19 classification tasks by extracting and fusing distant features. There are two main improvements made by our algorithm. Firstly, it does not require any labeled source domain data, and the source domains can be completely different from the target domain. The proposed model only needs a small amount of labeled target domain and can produce very promising classification accuracy on the target domain. Secondly, it only focuses on improving the performance of the target task in the target domain. To the best of our knowledge, it is the first time that DDTL has been applied to medical image classification. Furthermore, we introduce a novel feature selection method, Distant Feature Fusion (DFF), to discover general features across distant domains and tasks by using convolutional autoencoders with a domain distance measurement. To outline, there are four main contributions made in this study: 1) Propose a new DDTL algorithm for fast and accurate COVID-19 diagnose based on lung CT, 2) Examine existing deep learning models (transfer and non-transfer) on COVID-19 classification problem, 3) The proposed algorithms has achieved the highest accuracy on this task, which has a small set of labeled target data and some unlabeled source data from different domains. Moreover, compared with other transfer learning methods, supervised learning methods, and existing DDTL methods, the proposed DFF model has achieved up to 34\% higher classification accuracy and 4) The proposed algorithms can be easily generalized to other medical image processing problems.

The remainder of this paper is structured as follows: In Section~\ref{RW}, we first review the most recent DTTL works. And then, we formulate the problem definition in Section~\ref{PS}. Next, we present the details of the proposed algorithm in Section~\ref{ME}. After that, we present experimental results and analysis in Section~\ref{EA}. Lastly, we conclude the paper and discuss future directions in Section~\ref{CR}.

\section{Related Work} \label{RW}

Insufficient training data and domain distribution mismatch have become the two most difficult challenges in the study of machine learning. To address these two issues, transfer learning has emerged a lot of attention due to its training efficiency and domain shift robustness. However, transfer learning also suffers from a critical shortcoming, negative transfer \cite{ge2014handling}, which significantly limits the use and performance of transfer learning. In this section, we introduce some related works in three fields: conventional transfer learning, DDTL, and existing ML methods for COVID-19 classification.

\subsection{Conventional Transfer Learning}
First of all, transfer learning aims to find and transfer the common knowledge in the source domain and the target domain. Furthermore, \cite{long2017deep,Niu2006:Transfer,Liu2020} expanded the use of transfer learning from traditional machine learning models to deep neural networks. Typically, there are two types of accessible transfer learning: feature-based and instance-based. Moreover, both types focus on closing the distribution distance between the source domain and the target domain. In instance-based algorithms, the goal is to discover source instances similar to target instances, so that highly unrelated source samples would be eliminated. Differently, feature-based algorithms aim to map source features and target features into a common feature space where the distribution mismatch is minimized. However, both of them naturally assume that the source domain and the target domain share a fairly strong connection. Unlike conventional transfer learning, our work can transfer knowledge between different domains and tasks that are not closely related. 

\subsection{DDTL}
Secondly, most DDTL algorithms are similar to multi-task learning \cite{zhang2016deep}, which also benefits from shared knowledge in multiple different but related domains. Generally, multi-task learning tends to improve the performance on all tasks. Differently, DDTL only focuses on using the knowledge in other domains to improve the performance of the target task in the target domain. Moreover, most previous studies of DDTL focus on instance-based methods and tend to take advantage of massive related source data. Firstly, \cite{tan2015transitive} introduced an instance-based algorithm, transitive transfer learning (TTL). It transfers knowledge between text data in the source domain and the image data in the target domain by using annotated image data as a bridge. However, TTL is highly case-dependent and unstable in performance. Similarly, \cite{tan2017distant} introduced another instance-based algorithm with a novel instance selection method, Selective Learning Algorithm (SLA). Moreover, SLA is used to select helpful instances from a number of unrelated intermediate domains to expand the source domain's volume. However, this algorithm can only handle binary classification problems. Furthermore, \cite{xie2016transfer} proposed another feature-based method to deal with scarce satellite image data. It predicts the poverty based on daytime satellite images by transferring knowledge learned from an object classification tasks with the aid of nighttime light intensity information as a bridge. However, this method relies heavily on a massive amount of labeled intermediate training data, which can be too expensive to apply. Unlike existing DDTL algorithms, our method benefits from multiple source domains without labeled data, and those source domains can have significant discrepancies. Furthermore, our method can also handle multi-class classification while consistently producing promising results.

\subsection{Machine Learning for COVID-19 Diagnosis}
Moreover, to overcome the shortage of COVID-19 testing toolkits, many efforts have been made to search for alternative solutions. \cite{article} shows that lung CT images can be used for diagnosis. Furthermore, several studies \cite{he2020sample,chen2020simple,barstugan2020coronavirus} introduce machine techniques to COVID-19 diagnosis, including but not limited to, convolutional neural networks (CNN), transfer learning, empirical modeling. However, most existing models suffer from a common shortcoming that is insufficient well-labeled training data. And transfer leanings methods can carry out fairly decent classifications, but they are still limited by the domain discrepancy between the source data and the target data.

\section{Problem Statement} \label{PS}

In this DDTL problem, we assume that the data of each target domain is not enough to train a robust model. And we have a number of unlabeled source domains denoted as \begin{math} \it{S} = \left\{(x_1^1, ..., x_1^{n_{S_1}}), ..., (x_{S_N}^1, ..., x_{S_N}^{n_{S_N}})\right\} \end{math}, where \( n\) and \( S_N\) represent the number of samples in each source domain and the number of source domains. Then we denote one or multiple labeled target domains as:
\begin{equation} 
\begin{split}
\it{T} = [(x_1^1, y_1^1), ..., (x_{1}^{n_{T_1}}, y_{1}^{n_{T_1}})], \\
..., [(x_{T_N}^1, y_{T_N}^1), ..., (x_{T_N}^{n_{T_N}}, y_{T_N}^{n_{T_N}})]
\end{split}
\end{equation}, 
where \( n\) and \( T_N\) represent the number of samples in each source domain and the number of source domains. Let \( P(x)\), \( P(y|x)\) be the marginal and the conditional distributions of a data set. In this DDTL problem, we have the following:

\begin{equation} 
P_{S_1-S_N}(x) \neq P_{T_1-T_N}, 
\end{equation}

\begin{equation} 
P_{T_1}(y|x) \neq P_{T_2}(y|x) \neq ... \neq P_{T_N}(y|x). 
\end{equation}

The main objective of the proposed work is to develop a model for the target domain with a minimal amount of labeled data by finding generic features from distant unlabeled source domain data. The motivation behind this study is that data in distant domains is usually seemingly unrelated in the instance-level but related in the feature-level.  However, the connection on the feature level from one distant domain can be too weak to be used to train an accurate model. As such, simply using one or two sets of source data is likely to fail in building the target model. Therefore, we leverage from multiple unlabeled distant source domains to obtain enough information for the target task.

\section{Methodology} \label{ME}

In this section, we introduce the proposed COVID-19 diagnose framework. Firstly, we present the reduced-size ResNet segmentation model. After that, we introduce the novel DDTL algorithm, Distant Feature Fusion (DFF). 

\subsection{Lung CT Segmentation by Reduced-size ResUnet}

\begin{figure}[htb] 
    \centering
    \includegraphics[width=0.5\textwidth]{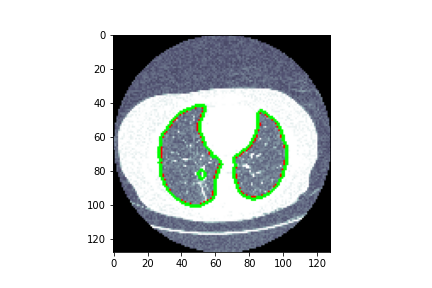}
    \captionsetup{justification=centering}
    \caption{Segmented Lung Area}
    \label{fig: SegImg}
\end{figure}

First of all, extracting features from a full size lung CT image with a small training set can be difficult because the model might end up focusing on noise in the useless parts of the images. Therefore, we propose to pre-process the image to improve the performance of the distant feature fusion by applying semantic segmentation. By doing this, as shown in Figure. \ref{fig: SegImg}, we can remove random noise and preserve the important information in the lung area of a image. Moreover, a small data set for training can lead to a over-fitting situation in a deep neural network. Therefore, we develop a reduced-size ResNet for this Covid-19 diagnose task. 

Similar to the original Unet \cite{ronneberger2015u}, the proposed reduced-size ResUnet contains two feature extraction parts: four convolutional blocks layers with down-sampling and four deconvolutional  layers with up-sampling. However, the lung-CT data set used for training is smaller than the data set used in the original paper. Therefore, we reduce the numbers of convolutional layers and deconvolutional layers, and apply dropout layers to prevent over-fitting. Furthermore, we adopt skip-connection to prevent two main problems in the training process: gradient explode and gradient disappear. In this study, we implement a single skip-connection to form convolutional and deconvolutional blocks. By doing this, the convergence time of the model is faster and the training process is more stable. 

Furthermore, image segmentation tasks require to perform accurate pixel-level classification on the input images. Therefore, it is critical to design a proper loss function based on each task. In this study, the final loss function is composed by a soft-max function over the last feature map combined with the cross-entropy loss. The expressions of the soft-max function and cross-entropy functions are:

\begin{equation} \label{SF}
p_k(x) = exp(f_k(x)) / \sum_{k=1}^{K} exp(f_k(x)),
\end{equation}

\begin{equation} \label{CE}
E = \sum_{x} \omega (x) log(p_{(l(x)})(x)),
\end{equation}
where $f_k(x)$ represents the activation map of the $kth$ feature at $xth$ pixel and $K$ is the total number of classes, and the cross-entropy penalizes at each position the deviation of $p_{(l(x)})$. Furthermore, the segmentation boarder is computed with morphological operations. The weight map is expressed as:

\begin{equation} \label{WM}
\omega (x) = \omega_c (x) \omega_0 exp(-\frac{(d_1(x) + d_2(x))^2}{2\sigma^2}),
\end{equation}
where $\omega_c$ is the weight map to balance the class frequencies, $d_1$ and $d_2$ are the distances between a pixel to the closest boarder and the second coolest boarder, and $\omega_0$ and $\sigma$ are the initialization values.

\subsection{DFF}

\begin{figure*}[htb]
    \centering
    \includegraphics[width=0.7\textwidth]{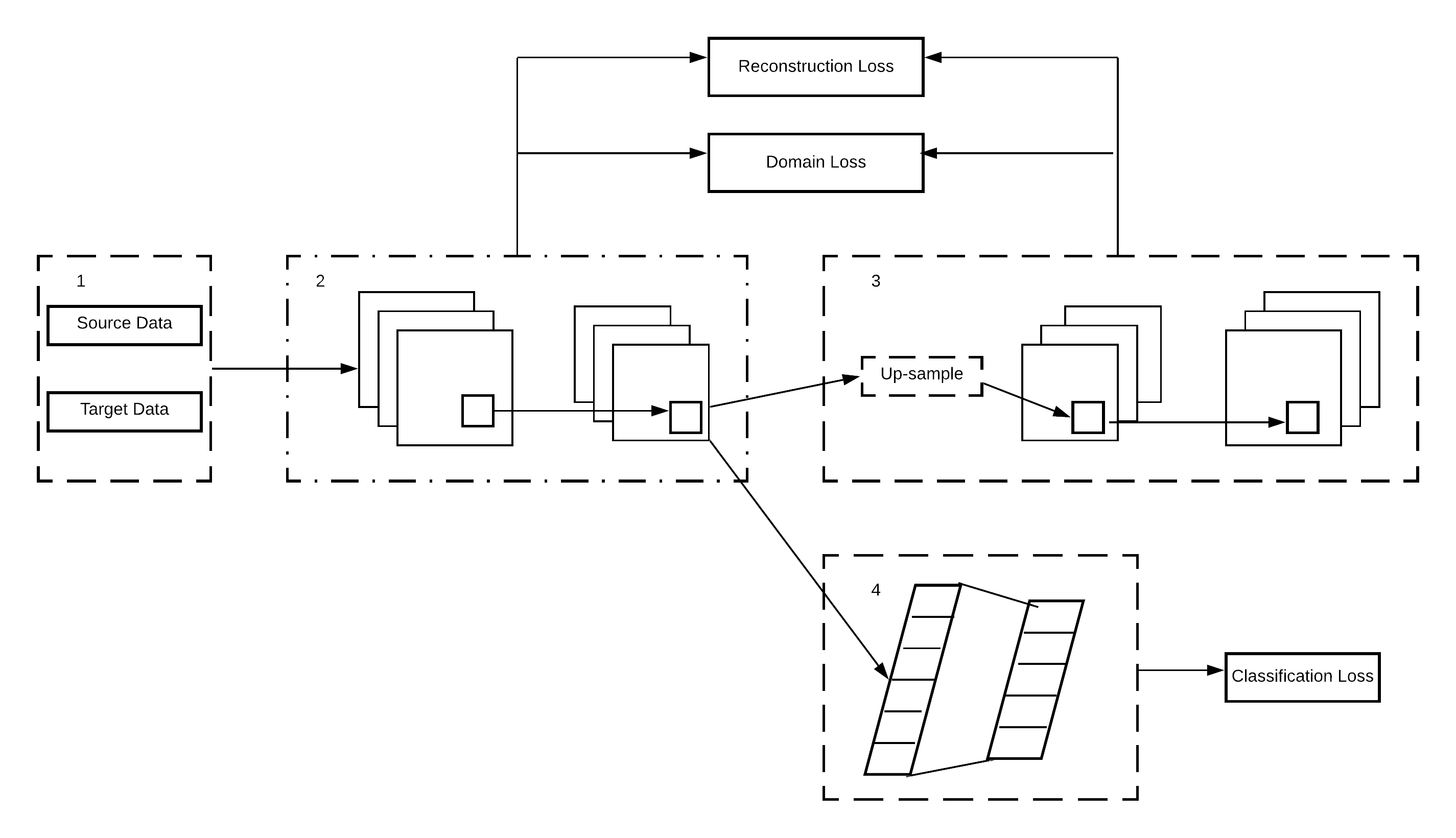}
    \captionsetup{justification=centering}
    \caption{DFF Architecture}
    \label{DFF}
\end{figure*}

As shown in Fig. \ref{DFF}, there are three main components in DFF: distant feature extractor, distant feature adaptation, and the target classification. There are three types of losses from three components: reconstruction loss, domain loss, and classification loss.

\subsubsection{Distant Feature Extraction}

As one of the inspirations of this study, a convolutional autoencoder pair is used as a feature extractor in DFF. As a variant of autoencoders, convolutional autoencoders \cite{turchenko2017deep} are usually beneficial for unsupervised image processing related problems. First of all, a convolutional autoencoder is a feed-forward neural network working in an unsupervised manner, which suits this DDTL problem perfectly since there is no labeled data in source domains. Generally, a convolutional autoencoder pair contains one input layer, one output layer, one up-sampling layer, and multiple convolutional layers. Moreover, there are two main components: encoder \( E_{Conv}(\cdot)\) and decoder \( D_{Conv}(\cdot)\). The standard process of convolutional autoencoder pairs can be demonstrated as:

\begin{equation} 
\textbf{\it{Encoding}}: f = E_{Conv}(x), \textbf{\it{Decoding}}: \it{\hat{x}} = D_{Conv}(\it{\hat{f}}),
\end{equation}
where \(f\) is the extracted features of \(x\), and \(\it{\hat{x}}\) is the reconstructed \(x\). In addition, the way to tune the parameters of a convolutional autoencoder pair is to minimize the reconstruction error on all the training instances. Conceptually, the output of the encoder can be considered as high-level features of the unlabeled training data. Furthermore, these features are learned in an unsupervised manner, so they are robust if the reconstruction error is lower than a certain threshold.

In this DDTL problem, as shown in Fig. \ref{ED}, we use a convolutional autoencoder pair to discover robust feature representation from unlabeled source domain data sets and the labeled target data sets simultaneously. And more, \(Module2\) and \(Module3\) are the encoder and the decoder. The structure of the auto-encoder pair contains two convolutional layers and two pooling layers in both the encoder and decoder. Up-sampling is applied to the encoder to ensure the quality of the reconstructed images. The process of feature selection has three main steps: feature extraction, instance reconstruction, and reconstruction measurement. First, we feed both the source data and the target data into the encoder to obtain high-level features \(f_S\) and \(f_T\). Then, extracted features are sent into the decoder to get reconstructions, \(\hat{X_S}\) and \(\hat{X_T}\). The equations of the first two steps are expressed as:

\begin{equation} 
f_S = E_{Conv}(X_S), f_T = E_{Conv}(X_T);
\end{equation}

\begin{equation} 
\hat{X_S} = D_{Conv}(f_S), \hat{X_T} = D_{Conv}(f_T);
\end{equation}
where \(X_S\) and \(X_T\) are the source and the target samples, and \(f_S\) and \(f_T\) are the source and the target features. Finally, we define the reconstruction errors from both the source domains and the target domains as the loss function of the feature extractor, \(L_R\) as follow:

\begin{equation} \label{LR}
\begin{split}
L_R = \sum_{i=1}^{S_N} \sum_{j=1}^{n_{S_i}} \frac{1}{n_{S_i}} (\hat{X_{X_{S_i}}^j} - X_{X_{S_i}}^j)^2 + \\
 \sum_{i=1}^{S_T} \sum_{j=1}^{n_{T_i}} \frac{1}{n_{T_i}} (\hat{X_{X_{T_i}}^j} - X_{X_{T_i}}^j)^2.
\end{split}
\end{equation}
where \(S_N\) and \(S_T\) are the numbers of the source domains and the target domains, \(n_{S_i}\) and \(n_{S_i}\) are the numbers of instances in the \(ith\) source domain and the target domain.

\subsubsection{Distant Feature Adaptation}

\SetKwInput{kwInit}{Initialize}
\begin{algorithm}[hbt] 
\SetAlgoLined
\KwIn{\(S = {X_S}\), \(T = {X_T, Y_T}\). \linebreak{Max Iteration: I, Batch Number: N.}}

\For {\(i = 1,...., I\)}
{

\For {\(j = 1,...., N\)}
{
Feature Extraction: \(f_S = E_{Conv}(X_S) \), \(f_T = E_{Conv}(X_T)\)
\linebreak{Instance Reconstruction: \(\hat{X_S} = D_{Conv}(X_S)\), \(\hat{X_T} = D_{Conv}(X_S)\)}
\linebreak{Label Prediction: \(X_{Pred}^T = C_T(f_T)\)}
\linebreak{Calculate \(L_R, L_D, L_C\)}
\linebreak{Update \(\theta_E,\theta_D,\Theta_C\)}
}
}
\KwOut{\(X_{Pred}^T\)}
\caption{Distant Feature Fusion Algorithm}
\label{Alg: DFF}
\end{algorithm}

Commonly, minimizing the reconstruction error \(L_R\) can discover a set of high-level features of the given input data. However, the distribution mismatch between the source and the target domains is significant, so minimizing\(L_R\) alone is insufficient to extract robust and domain-invariant features. Therefore, we need extra side information to close the domain distance, so the extracted features are robust to both the source domains and the target domains. In this research, as shown in Fig. \ref{DFF}, we add a distant feature adaptation layer to the convolutional autoencoder pair to measure the domain loss, \(L_D\). The maximum mean discrepancy (MMD) \cite{borgwardt2006integrating}, an important statistical domain distance estimator, is used as the domain distance measurement metric. The domain loss is expressed as:

\begin{equation} \label{LD}
L_D = MMD (\sum_{i=1}^{S_N} \sum_{j=1}^{n_{S_i}} f_{S_i}^j, \sum_{i=1}^{S_T} \sum_{j=1}^{n_{T_i}} f_{T_i}^j),
\end{equation}

\begin{equation} \label{MMD}
MMD (X, Y) = 	\parallel \frac{1}{n_1} \sum_{i=1}^{n_1} \varphi(x_i) + \frac{1}{n_2} \sum_{f=1}^{n_2} \varphi(y_j) \parallel,
\end{equation}
where \(n_1\) and \(n_2\) are the numbers of instances of two different domains, and \(\varphi(\cdot)\) is the kernel that converts two sets of features to a common reproducing kernel Hilbert space (RKHS) where the distance of two domains is maximized.

\subsubsection{Target Classifier}

Furthermore, with extracted high-level features, we add two fully-connected layers after the encoder to build a target classifier, \(C_T\), for the target task in the target domain. As the motivation of this step, \cite{krizhevsky2012imagenet} proves that convolutional layers can discover features, and fully-connected layers can find the best feature combination for each class in the target task. In other words, fully-connected layers do not learn more new features but connect each class to a specific set of features with different weights. In this work, there is only one fully-connected layer followed by the output layer with cross-entropy loss, \(L_C\):

\begin{equation} \label{LC}
L_C = -x[Class] + \sum_{i=1}^{T_N} \sum_{j=1}^{n_{T_i}} exp(X_{T_i}^j).
\end{equation}
where \(X_{T_i}^j\) is the \(jth\) sample in the \(ith\) target domain. Finally, by embedding all three losses from~\ref{LR},~\ref{LD}, and~\ref{LC}, the overall objective function of DFF is formulated as:

\begin{equation} \label{Ojb}
\underset{\theta_E,\theta_D,\Theta_C}{\text{Minimize}} \quad L = L_R + L_D + L_C,
\end{equation}
where \(\theta_E,\theta_D,\Theta_C\) are the parameters of the encoder, decoder, and the classifier, respectively. Moreover, \(L\) is the final loss constructed by the reconstruction error, domain loss, and classification loss. Finally, all the parameters are optimized by minimizing the objective function in Equation~\ref{Ojb}.

\subsubsection{Algorithm Summary}

Lastly, an overview of the proposed work is summarized in Algorithm~\ref{Alg: DFF}.

\section{Experiment and Analysis} \label{EA}

\begin{table*} 
  \caption{Model Comparison}
  \label{Tlb: MC}
  \centering
  \begin{tabular}{llllllll}
   \toprule
    \cmidrule(r){1-1}
          & CNN & Alexnet  & Resnet & SelfTran  & SLA & DFF \\
    \midrule
    Transferable & No & Yes  & Yes& Yes  & Yes & Yes    \\
    Base Model    & Discriminative & Discriminative & Discriminative  & Discriminative  & Discriminative & Discriminative         \\
    Loss Type     & Entropy  & Entropy & Entropy & Entropy  & Entropy\&MMD & Entropy\&MMD \\
    Learning Type     & Feature-based & Feature-based & Feature-based & Feature-based  & Instance-based & Feature-based   \\
    \bottomrule
  \end{tabular}
\end{table*}

In this section, we introduce a number of benchmark models, such as supervised learning models, conventional transfer learning models, and DDTL models. Then we set up several different combinations of the source and the target to conduct a number of experiments. Moreover, we represent results from the proposed models and comparisons with benchmark models and other TL models. Finally, we present training details and the analysis of experimental results. 

\subsection{Benchmark Models} 

\begin{table*} 
  \caption{Data Sets}
  \label{Tlb: DT}
  \centering
  \begin{tabular}{lllll}
   \toprule
    \cmidrule(r){1-1}
         Data Set & Total Classes & Total Samples & Label & Mask   \\
    \midrule
        Catech-256 & \(256\) & \(30670\) & \(Yes\) & \(No\)\\
        Office-31 & \(31\) & \(4110\) & \(Yes\) & \(No\)\\
        Chest Xray & \(4\) & \(562\) & \(Yes\) & \(No\)\\
        Lung-CT & \(4\) & \(367\) & \(Yes\) & \(Yes\)\\
        Covid19-CT & \(2\) & \(565\) & \(Yes\) & \(No\)\\
    
    \bottomrule
  \end{tabular}
\end{table*}

\begin{table} 
  \caption{Segmentation Performance}
  \label{Tlb: Seg}
  \centering
  \begin{tabular}{llll}
   \toprule
    \cmidrule(r){1-1}
          & IoU & Dice & Accuracy  \\
    \midrule
        Reduced-ResUnet & \(0.96\) & \(0.97\) & \(0.96\)\\
        Unet & \(0.86\) & \(0.88\) & \(0.87\) \\
    \bottomrule
  \end{tabular}
\end{table}

In this study, as shown in Table. \ref{Tlb: MC}, we choose several common transfer models and non-transfer models for comparisons. By comparing results from different methods, we can justify the improvements made by the proposed methods. Firstly, we select three supervised non-transfer baseline models: convolutional neural works (CNN), Alexnet \cite{krizhevsky2012imagenet}, and Resnet \cite{7780459}. For CNN, the model is constructed with three convolutional layers with \(3 \times 3\) kernels followed by a \(2 \times 2\) max polling kernel. Secondly, we also choose three conventional transfer learning models: fine-tuned Alexnet, fine-tuned Resnet, and self-transfer (SelfTran) model \cite{he2020sample}. Moreover, the convolutional layers of the Alexnet model and Resnet model remain frozen during the fine-tuning, and only the fully-connected layers are re-trained. What is more, we choose one instance-based DDTL method: selective learning algorithm (SLA) \cite{tan2017distant}. However, this DDTL model cannot be completely reproduced with many details not being introduced in the papers, and there is no accessible source code. As such, reproduced accuracy of this algorithm can be potentially improved with more careful tuning. Finally, we produce multiple experiments on the proposed DFF and ADFE algorithms. Furthermore, all details of each benchmark model are specified in Table. \ref{Tlb: MC}.

\subsection{Date Sets and Experiment Setups}

\begin{figure*}[htb]
    \centering
    \includegraphics[width=0.7\textwidth]{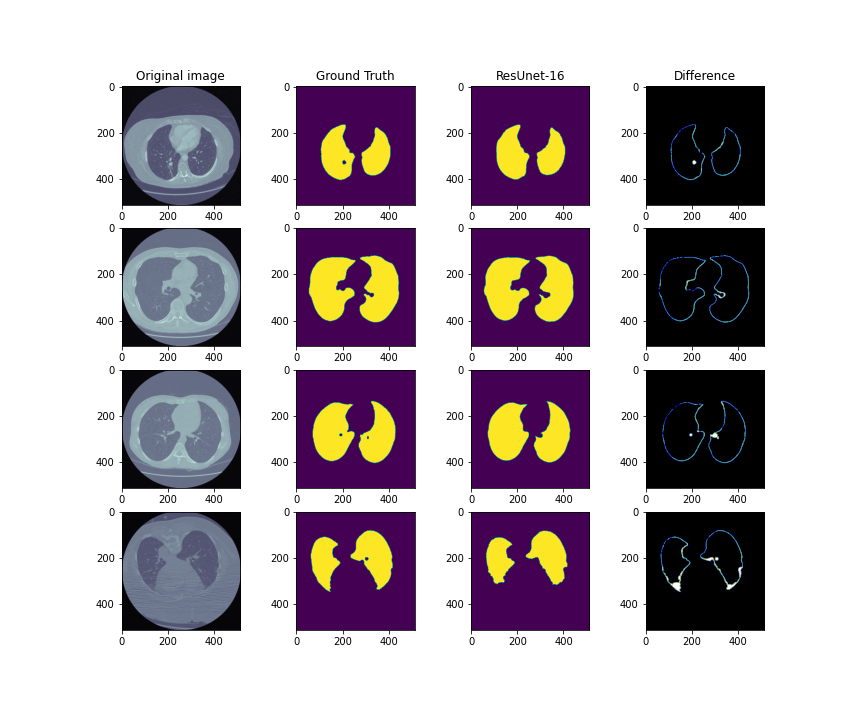}
    \captionsetup{justification=centering}
    \caption{Lung CT Segmentation}
    \label{fig: lungct}
\end{figure*}

In this study, as shown in Table. \ref{Tlb: DT}, we totally use six open-source data sets: Catech-256 \cite{griffin2007caltech}, Office-31 \cite{Zhao2017StretchingDA}, chest X-Ray for pneumonia detection \cite{kermany2018identifying}, Lung CT \cite{armato2011lung}, and Covid19-CT \cite{zhao2020COVID-CT-Dataset}. The first, Catech-256 is an image data set that includes labeled data of 256 different classes. For each class, the number of instances is from 80 to 827. Then, Office-31 has three collections of 31 different common office objects, with total 4110 instances collected from three different data sources: "amazon", "webcam", and "dslr". However, Office-31 is acknowledged as an extremely unbalanced data set because the numbers of samples from the three domains are very different. Moreover, the chest X-Ray data set contains 5226 well-labeled images. Intuitively, the chest X-Ray images should have the most similarity with lung X-Ray images, so we wonder if directly transfer and fine-tune on this data set would carry out better performance than the proposed methods. Moreover, Covid19-CT contains 565 labeled lung CT images: 349 positive samples, and 216 negative samples. It is considered as a fairly small data set for training deep learning models. Finally, we use the lung CT data set for the segmentation model. The data set has 367 lung CT images with pixel-level masks.

\begin{table*} 
  \caption{Top Accuracies (\%) of Examined Models}
  \label{Tlb: TA}
  \centering
  \begin{tabular}{lllllll}
   \toprule
    \cmidrule(r){1-1}
          & CNN & Alexnet & Resnet & SelfTran  & SLA & DFF  \\
    \midrule
        Testing Accuracy (Raw-Image) & \(74\pm{1}\) & \(82\pm{3}\) & \(86\pm{3}\) & \(83\pm{1}\) & \(54\pm{2}\) & \(\bf{93}\pm{1}\)   \\
        Testing Accuracy (Segmented-Image) & \(78\pm{2}\) & \(85\pm{2}\) & \(88\pm{3}\) & \(87\pm{3}\) & \(62\pm{1}\) & \(\bf{96}\pm{1}\)   \\
    \bottomrule
  \end{tabular}
\end{table*}

\begin{table*} 
  \caption{Accuracies (\%) of DDTL Models with Single Source Domain}
  \label{Tlb: DS}
  \centering
  \begin{tabular}{llllll}
   \toprule
    \cmidrule(r){1-1}
    DDTL Models &&&&&\\
     \midrule
       Source Domain   & Catech256  & Amazon  & Webcam  & Dslr & Chest X-Ray \\
    \midrule
     SLA (Raw-Image) & \(54\pm{2}\) & \(52\pm{1}\)  & \(48\pm{2}\)  & \(48\pm{3}\)  & \(52\pm{4}\)   \\
     SLA (Segmented-Image) & \(62\pm{1}\) & \(54\pm{1}\)  & \(46\pm{3}\)  & \(56\pm{1}\)  & \(61\pm{2}\)   \\
    DFF (Raw-Image) & \(\bf{88}\pm{2}\) & \(78\pm{3}\)  & \(73\pm{2}\) & \(70\pm{1}\)  & \(63\pm{3}\)  \\
    DFF (Segmented-Image) & \(\bf{90}\pm{1}\) & \(76\pm{1}\)  & \(76\pm{2}\) & \(74\pm{3}\)  & \(69\pm{2}\)  \\
    \midrule
    Conventional TL Models &&&&&\\
    \midrule
    Fine-tuned Alexnet (Raw-Image)   & \(77\pm{1}\) & \(61\pm{2}\)  & \(64\pm{1}\) & \(51\pm{2}\) & \(73\pm{3}\)  \\
    Fine-tuned Alexnet  (Segmented-Image)  & \(80\pm{2}\) & \(64\pm{1}\)  & \(68\pm{3}\) & \(52\pm{2}\) & \(81\pm{1}\)  \\
    Fine-tuned Resnet (Raw-Image) & \(66\pm{2}\) & \(57\pm{3}\)  & \(61\pm{1}\) & \(54\pm{1}\) & \(64\pm{2}\)  \\
    Fine-tuned Resnet (Segmented-Image)   & \(72\pm{1}\) & \(61\pm{1}\)  & \(64\pm{2}\) & \(62\pm{3}\) & \(65\pm{2}\)  \\
    \bottomrule
  \end{tabular}
\end{table*}

\begin{table*} 
      \caption{Accuracies (\%) of DDTL Models with Multiple Source Domains}
  \label{Tlb: DM}
  \centering
  \begin{tabular}{lccccl}
   \toprule
    \cmidrule(r){1-1}
      Primary Source Domain   & Catech256  & Amazon  & Webcam  & Dslr \\
    
    Auxiliary Source Domain  & \multicolumn{4}{c}{Chest X-Ray} \\
    \midrule
     SLA (Raw-Image) & \(54\pm{2}\) & \(52\pm{1}\)  & \(48\pm{2}\)  & \(48\pm{3}\)    \\
     SLA (Segmented-Image) & \(62\pm{1}\) & \(55\pm{3}\)  & \(51\pm{1}\)  & \(47\pm{2}\)    \\
    DFF (Raw-Image) & \(\bf{93}\pm{1}\) & \(73\pm{3}\)  & \(64\pm{2}\) & \(86\pm{3}\)   \\
    DFF (Segmented-Image) & \(\bf{96}\pm{1}\) & \(75\pm{2}\)  & \(66\pm{1}\) & \(87\pm{1}\)   \\
    \bottomrule
  \end{tabular}
\end{table*}

Moreover, we run each experiment five times to investigate the accuracy fluctuation range. Firstly, we produce 4 experiments on CNN and conventional TL models with the Covid19-CT data. In addition, Alexnet and SelfTran models are initialized with the parameter pre-trained on the ImageNet data set, and the CNN model is trained from scratch. And then, we set up a series of  experiments on DDTL models with single source domain and multi-source domains to explore the potential of the learning method. As shown in Table. \ref{Tlb: DS}, there are five unlabeled source domains data sets: \textbf{\textit{Catech-256}}, \textbf{\textit{Amazon}}, \textbf{\textit{Amazon}}, \textbf{\textit{Webcam}}, \textbf{\textit{Chest X-Ray}}, and one labeled target data set: \textbf{\textit{Lung CT for Covid-19}}. What is more, another regular \textbf{\textit{Lung CT}} contains masks for segmentation. Moreover, the first four source domains are seemingly unrelated to the target domain, but the last source domain is visually related to the target domain. 

Furthermore, unlike previous methods, the proposed method is able to utilize multiple source domains to improve the performance in the target domain. Therefore, as we can tell from Table. \ref{Tlb: DM}, we choose four primary source domains and use the \textbf{\textit{Chest X-Ray}} data set as the auxiliary domain. In the following sections, we will present the results and analysis. 

\subsection{Performance and Analysis}

In this section, we first present the performance of the segmentation model. After that, we give an overview of results of all examined classification methods and present insights on differences in performance. Then, we provide training details and analysis of our proposed DDTL algorithm.

\subsubsection{Segmentation Performance}

Firstly, the most informative part of a lung CT is the lung area, and it allows machines to better imitate the behaviors of real specialists. The proposed reduced-size ResUnet is trained from scratch because there is no pre-trained weights can fit this novel architecture. Moreover, the drop-out layers and the skip-connections are applied to prevent over-fitting and non-convergence problems. As we can tell from figure. \ref{fig: SegImg}, the segmented image shows an accurate and clear contour of the lung area, so we can select only the lung area as the input for the DFF model. Furthermore, Figure. \ref{fig: lungct} shows a better visual results of the segmentation model. The first column presents the original image, the second column shows the ground truth of the lung area, the third column gives the pixel-level classification of the model, and the fourth column illustrates the pixel-level difference between the ground truth and the prediction.

Moreover, we use two common evaluation metrics for image segmentation tasks to quantify the performance. In the study, we use IoU (intersection over union), Dice (F1 Score), and pixel-level accuracy as the evaluation metrics. The definitions of them are:

\begin{equation} \label{IOU}
IoU = \frac{TP}{TP + FP + FN},
\end{equation}

\begin{equation} \label{DICE}
Dice = \frac{2TP}{2TP + FP + FN},
\end{equation}

\begin{equation} \label{Accy}
Accuracy = \frac{TP + TN}{TP + TN + FP + FN}.
\end{equation}
Furthermore, for the comparison, we also conduct experiments on the original Unet with the same data set. The details are shown in Table. \ref{Tlb: Seg}. Obviously, the reduced-size ResUnet outperforms the original Unet. The possible reasons are: 1) the original Unet cannot effectively prevent the model learning from noise, 2) the skip-connections help the model to extract deeper features.

\subsubsection{Classification Performance Overview}

As demonstrated in Table. \ref{Tlb: TA}, the proposed DFF algorithm outperforms the highest test classification accuracy (\(96 \%\)). And more, the CNN model is only at (\(78 \%\)) classification accuracy. Intuitively, it is caused by insufficient training data. Moreover, the Alexnet and SelfTran result in promising accuracies (\(85 \%\), \(88 \%\)). In theory, initializing with pre-trained parameters can be the main reason to allow them achieve better because the [re-trained model was trained on a massive amount of images. However, in this case, the settings are more or less similar to TL, and the accuracies are still lower than the proposed DDTL method. This performance gap can be caused by large domain statistical discrepancy between two distant domains. In other words, the pretraining data sets have large domain distribution mismatches with the Covid19-CT data set, and the traditional models are not able to close the domain distance. Therefore, performance degradation cannot be avoided, but there is no evidence of any negative transfer in the fine-tuning models. Furthermore, an instance-based DDTL model (SLA) outputs the worst accuracy (\(62 \%\)), which is clearly a negative transfer case. The theoretical reason for it is that the instance selection by the re-weighting matrix ends up eliminating most source domain samples due to a large distribution discrepancy. As such, the selected instances are not enough to extract enough information for the knowledge transfer, which can be considered as the same situation as the CNN model with insufficient training data. Furthermore, it clearly shows that pre-processing the data with semantic segmentation can improve the performance. It justifies that preserving the most informative part by eliminating random noise from a Small data set can enhance the final classification performance.

Furthermore, there are other interesting things we can observe from the same table. First of all, feature-based algorithms have more promising performances in this COVID-19 classification problem. Compared with feature-based methods, the instance-based method completely failed to solve this task. Intuitively, samples in distant domains are seemingly unrelated at the instance level, but they might still share common information at the feature level. Therefore, the instance selection method can easily miss important information since it can only study data sets at the visual-level. Differently, the feature-based models tend to ignore the large discrepancy at the visual-level. Instead, they aim to discover the relationship of two domains at the feature-level and close the distribution mismatch by extracting domain-confusing features. 

\begin{figure*}[htb]
\centering  
\subfloat[Catech256-COVID19]
{
    \includegraphics[width=0.4\linewidth]{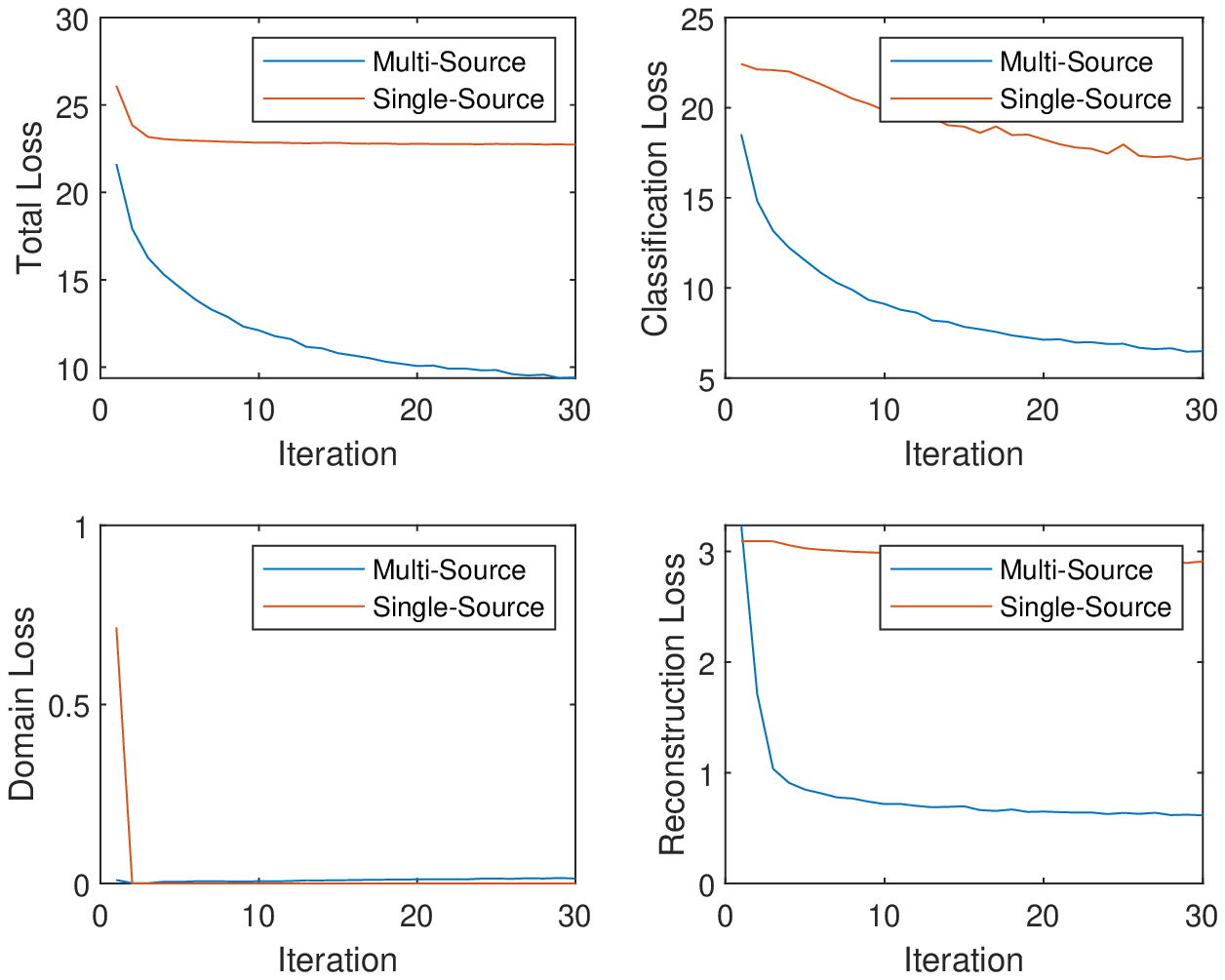}
    \label{1}
}
\subfloat[Amazon-COVID19]
{
    \includegraphics[width=0.4\linewidth]{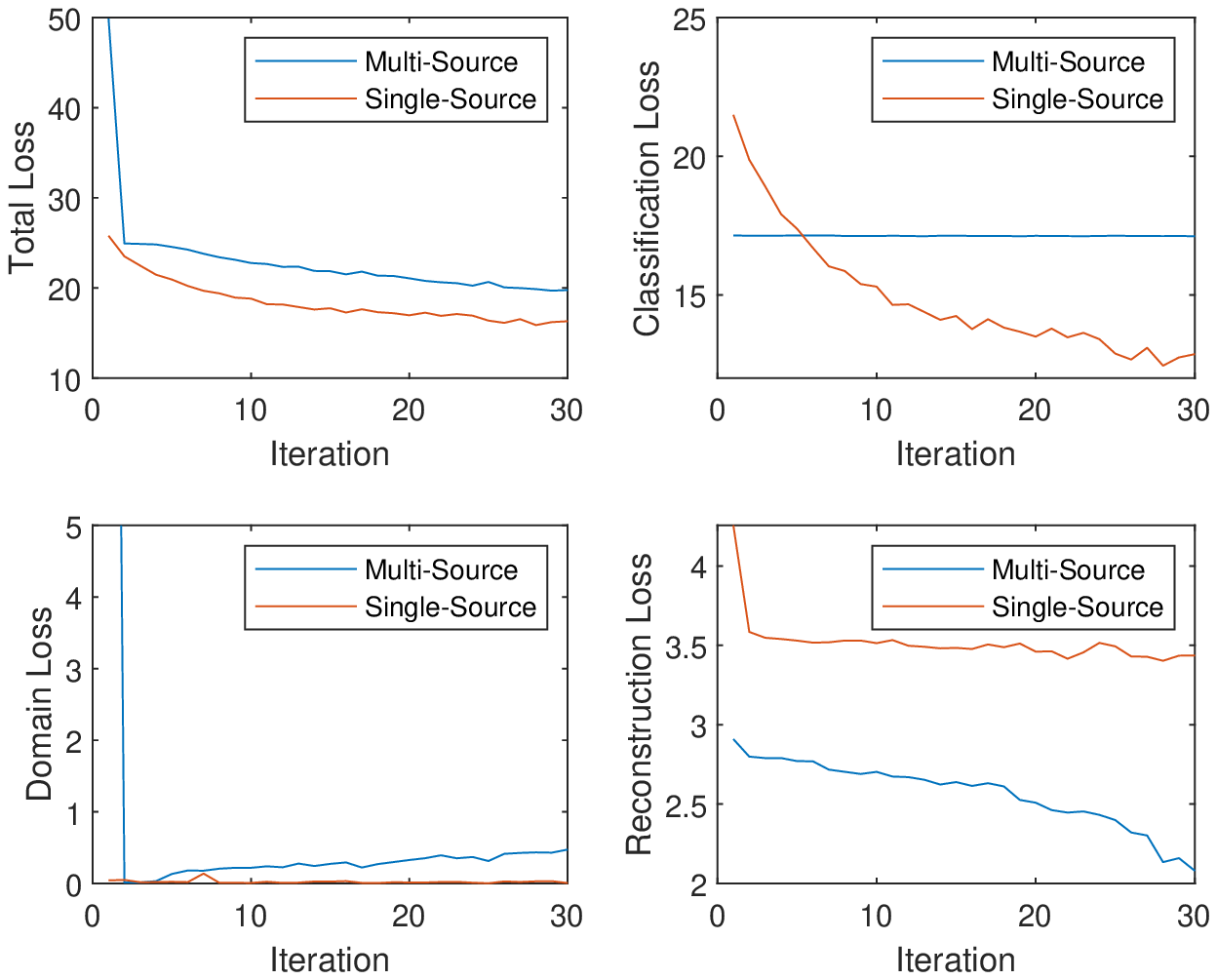}
    \label{2}
}\hspace*{-1.5em}\\
\subfloat[Webcam-COVID19]
{
    \includegraphics[width=0.4\linewidth]{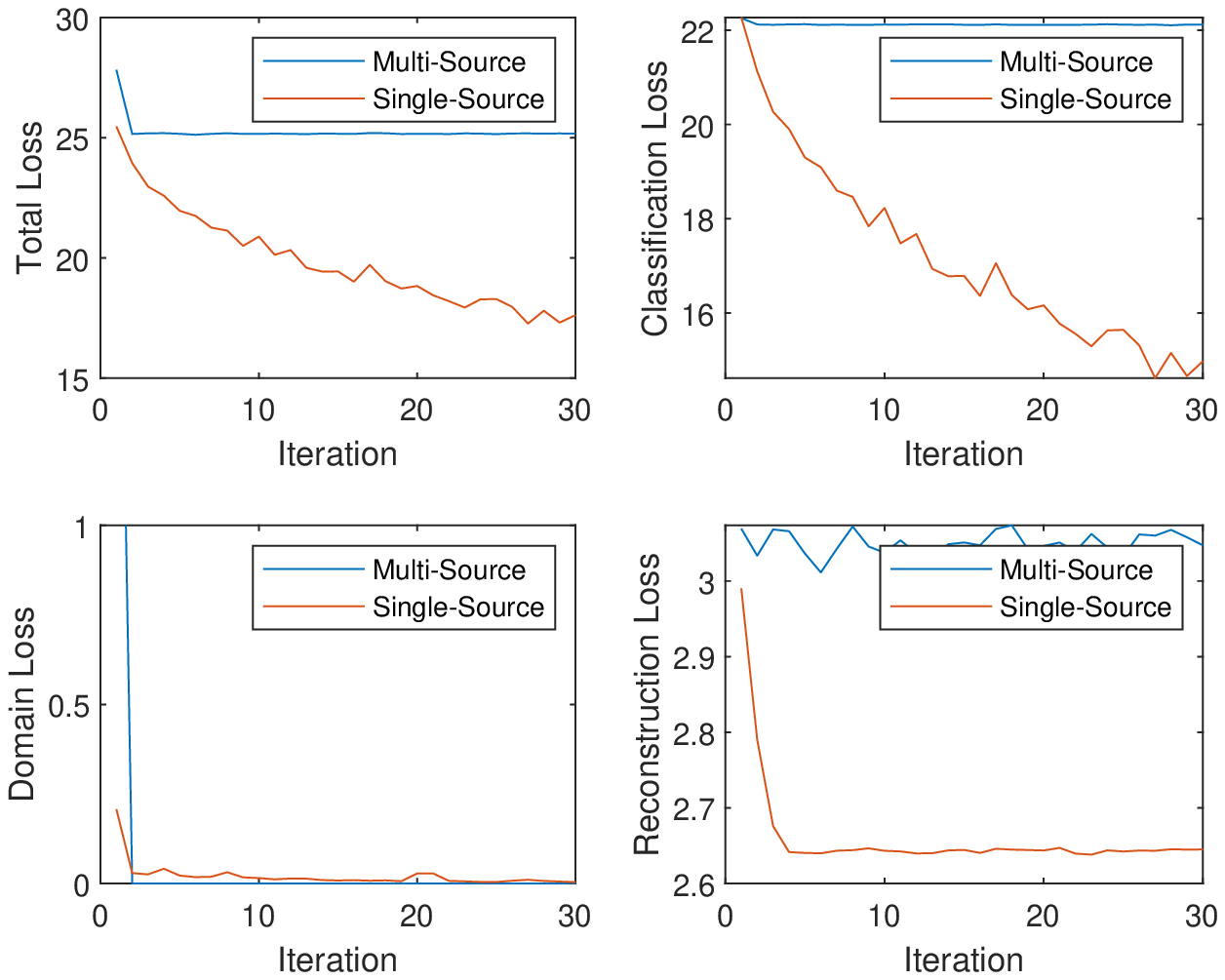}
    \label{3}
}
\subfloat[Dslr-COVID19]
{
    \includegraphics[width=0.4\linewidth]{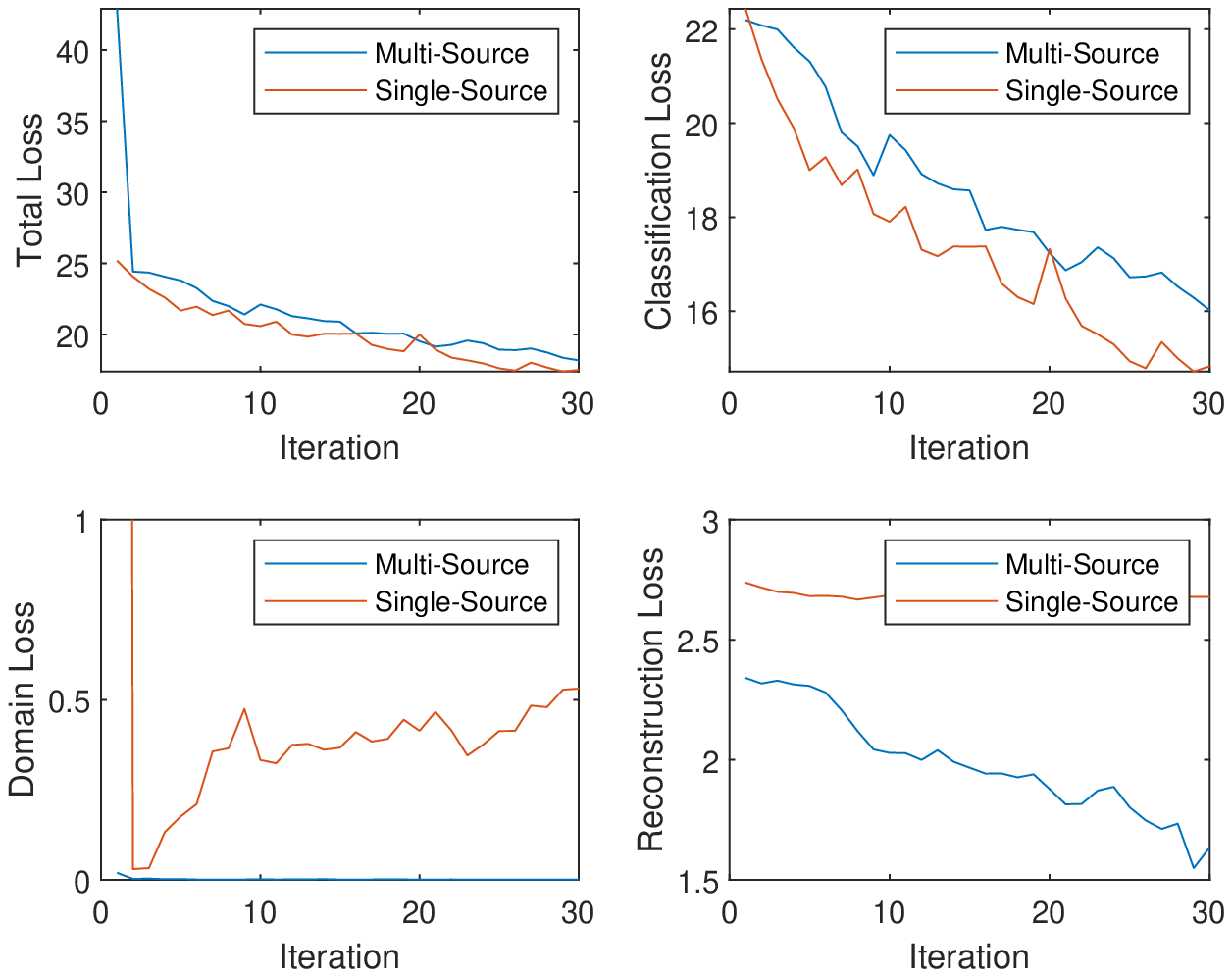}
    \label{4}
}\hspace*{-1.5em}

\caption{Training Details of experiments on ADFE with 4 setups: \textbf{\textit{Catech-256 to Covid19-CT}}, \textbf{\textit{Office-31-Amazon to Covid19-CT}}, \textbf{\textit{Office-31-Webcam to Covid19-CT}}, \textbf{\textit{Office-31-dslr to Covid19-CT}}. In each sub-figure, up left is total loss, up right is target classification loss, down left is domain distance, and down right is reconstruction error.}
\end{figure*}

Moreover, Table. \ref{Tlb: DS} shows performances of conventional TF models and DDTL models with single source domain. First of all, there are two completely different results from two DDTL methods. The proposed DDTL algorithm achieves the highest classification accuracy (\(90 \%\), but the previous instance-based SLA method occurs negative transfer on all five source domains. It further approves that instance selection process might carry out decent results when the source domain and the target domain have close relationships, but it does not suit the cases where two domains share a loose connection. Differently, feature-based methods are more reliable on DDTL problems. However, the advantage of SLA is that it does not require labeled target data, while the proposed method needs labeled target data. Moreover, not all source domains are suitable for distant knowledge transfer. The seemingly related domain, chest X-Ray, is actually not the most transfer-friendly for this task. Other data sets that are visually distant from the target domain carry out better results. It approves the theory that seemingly unrelated domains might be statistically connected in the feature-level. We will provide more evidences in later contents.

What is more, the best performance of conventional TL models is (\(88 \%\) which is better than non-transfer methods. Initializing with pre-trained weights only yields a faster convergence but it is not very helpful to improve the performance. Accuracies from experiments of \textbf{\textit{Chest X-Ray to Covid19-CT}} turns out to be worse than other experiment setups even the chest X-Ray is commonly assumed to be the most similar to the target domain. However, as shown in Fig. \ref{fig: DE}, the domain loss between the Covid19-Xray and chest X-Ray is the greatest among all experiments. Therefore, it also proves that seemingly related domains might be distant in the feature level, so it is not always reliable to hand-pick source domains in DDTL problems.

Moreover, the enhancement from semantic segmentation is still not good enough to reach the human-level performance. Therefore, unlike most existing DDTL algorithms, we wish to even improve the performance by using more than one source domain. Importantly, in DDTL problems, finding shared information cross different domains is the key to perform a safe knowledge transfer. However, the amount of common information extracted from a single distant domain might not be sufficient. As shown in Table. \ref{Tlb: DM}, the proposed method achieves (\(96 \%\) classification accuracy with using \textbf{\textit{Catech-256}} as the primary source domain and \textbf{\textit{Chest X-Ray}} as the auxiliary source domain. It means that these two data sets have less information overlapping, so the DFF model can extract more useful shared knowledge to transfer to the target domain. Differently, performance degradation appears in others multi-source domain experiments, which means others pairs have shared information that causes over-fitting.  

However, one significant weakness of DDTL models is that they are highly dependent on the quantity and versatility of the source domains. As we can tell from Table. \ref{Tlb: DS}, the performances of the DFF model and the ADFE model decrease dramatically when the webcam and the dslr data sets of Office-31 are set as the source domains. The cause of the performance drop is simple. Theoretically, DDTL models benefit from extracting the common knowledge of the source domain and the target domain, but they cannot complete this type of feature extraction when the source data set is small. There are only 550 and 640 samples in the webcam and the Dslr data sets, which are less than the target samples. Therefore, it is not easy to safely and effectively transfer knowledge between different domains. On the contrary, the Catech-31 data set has over 33000 samples from 256 different classes, so it is easier to perform the knowledge transfer.

\subsubsection{Analysis of DFF}

\begin{figure}[htb]
    \centering
    \includegraphics[width=0.4\textwidth]{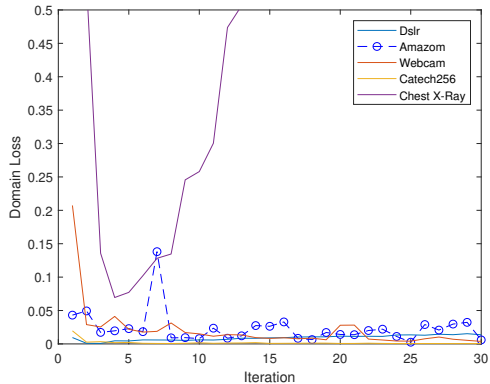}
    \captionsetup{justification=centering}
    \caption{DFF Domain Losses with Single Source Domain}
    \label{fig: DE}
\end{figure}

Fig. \ref{1}-\ref{4} shows details of the DDF models in single source domain setting and the multi-source domain settings, illustrating four types of losses: total training loss, target classification loss, domain loss, and reconstruction loss. Firstly, The proposed DFF algorithm has achieved the highest test classification accuracy when the Catech-256 data set is chosen as the primary source domain and the chest X-Ray data set as the auxiliary source domain. Overall, it has the most smooth curves and the smallest domain loss. Moreover, with the additional information from the auxiliary source domain, its classification loss and reconstruction loss are dramatically reduced. In other words, the model is able to extract additional features from the auxiliary domain, and use it as a bridge to close the distance from the target domain. Moreover, large declines in performance appear in the other experiments with \textbf{\textit{Amazon}} and \textbf{\textit{Webcam}}. As mentioned earlier, the performance degradation can be caused by overlapping information in the primary and the secondary source domains. The model tends to over-fit on the repeated knowledge in two source domains. Especially, in the experiment \ref{2}, the domain loss is increased and the classification loss can not be lowered. Furthermore, this provides another approve that seemingly distant instances might share a certain amount of common features, and such features can be extracted by properly adding a domain loss to the loss function. Moreover, Fig. \ref{fig: DE} supports another point: the smaller domain loss means a closer distance between two domains. As we can tell from the figure, the \textbf{\textit{Catech-256 to Covid19-CT}} combination has the lowest domain loss, and it also has the best classification accuracy. Furthermore, the domain loss curve of Dslr data set increases during the training, indicating that the quantity and the versatility of the source data set play an important role in this task. Finally, we quantify the performance of DFF model with four evaluation metrics: accuracy, precision, recall, and F1 score.

\begin{table*} 
  \caption{DFF Performance}
  \label{Tlb: DFFP}
  \centering
  \begin{tabular}{lllll}
   \toprule
    \cmidrule(r){1-1}
        DFF  & Accuracy & Precision & Recall & F1  \\
    \midrule
        Single Source & \(0.86\) & \(0.92\) & \(0.86\) & \(0.88\)\\
        Multi-Source & \(0.88\) & \(0.92\) & \(0.93\) & \(0.92\) \\
       Segmented Multi-Source & \(0.96\) & \(0.97\) & \(0.98\) & \(0.97\) \\
    \bottomrule
  \end{tabular}
\end{table*}

\section{Conclusion \& Future Work} \label{CR}

To draw a conclusion, in this paper, we introduce a novel DDTL algorithms, DFF, for a COVID-19 diagnostic method based on lung CT images. To distinguish from all conventional TL algorithms, the propose methods can use seemingly unrelated data sets to develop an efficient classification model for COVID-19 diagnose. Unlike previous DDTL models, our method enables knowledge transfer from multiple distant source domains, and it can effectively enhance the performance. Moreover, the proposed methods greatly expand the usage of transfer learning on medical image processing by safely transferring the knowledge in distant source domains, which can be completely different from the target domain. Moreover, this study is related to one of the most challenging problems in transfer learning, negative transfer. To the best of our knowledge, this is the first study that uses distant domain source data for COVID-19 diagnosis and outperforms promising test classification accuracy. Four contributions of this paper are made: 1) it successfully adopts DDTL methods to COVID-19 diagnosis, 2) we introduce a novel feature-based DDTL classification algorithms, 3) the proposed methods achieve state-of-art results, and 4) proposed methods can be easily expanded to other medical image processing problems. 

However, there are several drawbacks of DDTL algorithms: 1) most algorithms tend to be case-specific, 2) source domain selection is too complicated in some cases, 3) distant feature extraction process is computationally expensive.

In the future, there are a number of research directions regarding COVID-19 diagnosis and DDTL problems. Firstly, the explainability of the feature-based DDTL algorithm is a challenging but essential topic. Visualizing the changes on features in deep layers through the training process can not only help us to better understand the domain adaptation in the feature level and decision making process of deep ANN models, but also discover the relationship between two distant domains. Moreover, how to improve the efficiency of feature extraction process is another key to improve the performance. Commonly, generative adversarial networks (GANs) is widely acknowledged as a better feature extraction method. However, how to avoid non-convergence in the training process of adversarial networks is very challenging, and gradient explode and disappear make the training process for adversarial networks extremely difficult. As an inspiration, designing new adversarial loss functions is a possible way of dealing with this problem. Additionally, cross-modality TL, such as from image to audio, can be another potential solution to DDTL problem since semantic information can also exist in different cross-modality domains. Solving this problem can expand the use of transfer learning to an even higher level. Furthermore, for multi-source DDTL algorithms, source domain selection is important to stabilize the performance. Recently, active learning methods attract more and more attention from researchers. Finally, using medical CT images from other diseases as the source domain might or might be able to produce better results because seemingly related domains can also have large discrepancies in the feature level. Moreover, image data sets are usually not easy to access, so it is not always feasible to develop a TL model by using medical image data from other diseases. Therefore, granting access to medical image data sets to the public and generating distribution shift embedded artificial data is a promising future research direction in the field of medical image processing.

\bibliography{Ref}
\bibliographystyle{IEEEtran}
\end{document}